# Discussion of digital gaming's impact on players' well-being during the COVID-19 lockdown

Dr. Hiroko Oe
Bournemouth University
Faculty of Management
89 Holdenhurst road
Bournemouth BH8 8EB, UK
hoe@bournemouth.ac.uk

**Abstract**

This research discusses how to utilise digital gaming to support the well-being of its users and sustain their physical and mental health during the COVID-19 lockdown, in which people's activities are limited. The published academic literature that is written in English and available for access on online databases was reviewed to develop key take-aways and a framework for discussing how to enhance people's well-being in the COVID-19 lockdown. Interaction with other players in virtual communities has been found to have a positive influence on the mental health of those suffering from a lack of societal connection. A framework for further research has also been developed that focuses on the critical situation of the COVID-19 lockdown, as this is an urgent topic with a huge impact on our health. Some gaming service providers have been proactive in redesigning games' programming to be suitable for the lockdown situation, and this enables players to enjoy physical activities even at home.

1. Introduction

Augmented reality (AR) games such as Pokémon GO are very popular, with 850 million downloads in the greater city of Tokyo. It has been said that these games are making a real difference to people's lives by getting them out and about. This is just one example of how 'gamification' is being intertwined with our well-being and overall health. With ongoing development based on the available technologies, the positive aspects of gaming have the potential to support our health. This has increased significance during the COVID-19 lockdown, and it is a critical theme for researchers as it can help to sustain people's health while other activities are limited to avoid the spread of the virus.

1.1 Games and technologies

There are various resources linked with contemporary art, animation, games, AR and virtual reality (VR), and there is the potential to collaborate these resources and apply the associated technologies to a new phase of games in the context of well-being. For instance, the Internet of Senses (IoS) encompasses the use of advanced technologies, such as AR and VR, in the virtual interactive community. These computing technologies and the content that flows on their platforms enable people to experience a world beyond reality (Lichty, 2019). This also applies to the world of gaming. By 2030, technological advances will have moved humans away from a screen-based world and into a sense-based world (Bayern, 2019).

The enhancement of connected sustainability could have a positive impact on human beings. The IoS will enable contemporary experiences, but more importantly, it also has the potential to push forward the notion of connected sustainability (Kelly, 2020) and allow people to feel connected with other people (Wulf et al., 2020). While empirical and experimental case studies remain at a nascent stage of development (e.g. eScent), there is a need to start conversations with the major players in IoS ecosystems (businesses, citizens' groups, schools and relevant regulators) to boost the impact of gaming on people's health (Racat & Capelli, 2020).

1.2 Research gap

The technological advancement of gamification with VR has provided new topics for research. There have been discussions of both the pros and cons of the impact of digital gaming. For instance, Hawi et al. (2019) discussed the negative impact of gaming in their investigations on the associated addictive behaviour, whereas Watanabe et al. (2017) and Cheng (2019) discussed the positive impact of digital gaming in reducing stress and contributing to peoples' well-being. The trend of recent discussions indicates that researchers have developed some practical ways to measure how gaming contributes to the mental health of gamers (Ghazali et al., 2019; Hamari et al., 2019; Oh et al., 2018). Moreover, Vella et al. (2019) discussed the impact of gaming on enhancing social connectivity, and Chew and Mitchell (2019) suggested that gaming could be a positive catalyst in enhancing players' life experiences through interactions with other gamers in the virtual community. However, detailed discussions on digital gaming's impact on people's well-being that result in contributions to helping researchers and practitioners to design health-related schemes and propositions are required. The present study will focus on this theme, specifically while under the stress of COVID-19 lockdown.

1.3 Aims and objectives

This study aims to explore the impact of digital gaming on people's happiness and well-being, particularly during the COVID-19 lockdown situation. To achieve this, three objectives have emerged:

1. To critically review relevant academic discussions in the field of digital gaming and its impact on players' well-being;

2. To develop research questions with sub-themes to propose a discussion framework for further study;

3. To discuss the contribution of this study and present further research opportunities that could contribute to enhancing players' well-being in the COVID-19 lockdown situation.

2. Literature review: development of key topics for the study

2.1 Gamification: lockdown conditions of COVID-19

Recently, the positive impact of gaming on people has been concentrated with the theme of well-being and activities that can increase our health. However, the original concept of 'gamification' was discussed with a huge expectation in the business context. For instance, at the Serious Game Summit during the 2011 Game Developers Conference, lecturer Jane McGonigal provided her insights on the growing industry trend to emphasise the potential of gamification to enhance new horizons for businesses (McGonigal, 2020). Another lecturer, Dr Amy Jo Kim, a neuroscientist and entrepreneur, said that people must reconsider how they apply this concept to real social contexts (Xu, 2015). Dr Kim was concerned that the word 'gamification' had not been embedded in real societal and economic scenarios and could not produce active suggestions for solving social and economic issues. At this point in time, she had already pointed out that the potential of gamification should be stretched further in various societal settings to solve issues and enhance its contributions.

Since people were first put under the pressure of lockdown because of COVID-19, at the end of March 2020, Niantic, the launcher of Pokémon GO, has already established a number of ways to help people cope with the current situation, in which players cannot play as a group outdoors. For instance, new gameplay forms have been introduced that enable players to enjoy interactive activities, such as Go Battle League, from home by adjusting the requirements of walking distance from 3km to 0km. The company also established incentives, such as giving a discount on monster balls so that players can catch Pokémon without having to walk around outdoors. By providing more content and increasing the number of players' virtual possessions, players indoors can give and receive more gifts, which can be an incentive to play (Takahashi, 2020).

2.2 Contribution to well-being by walking

Around 2015, another wave arrived with a new perspective suggesting that gamification can contribute to good health (Johnson et al, 2016; Reer & Quandt, 2020). Although it is generally agreed that physical exercise is one of the key ways to maintain good health, public sector intervention aimed at increasing the take-up of exercise in cities has not been successful (NHS, 2019; Watanabe et al., 2017).

From this, a theme emerged of how to fuse the critical factors of digital gaming and well-being, particularly during the COVID-19 lockdown. After the first three weeks of lockdown, the rules were extended for at least another three weeks (as of 16 April 2020) (BBC News, 2020). The rules include restrictions on going outside. Only limited exercise is allowed outdoors: one form of exercise a day, for example a run, walk, or cycle – alone or with members of your household (Cabinet Office, 2020). This physical exercise must be conducted as an individual or with other people who live in the same house, so people are not allowed to hike or picnic with a group of friends.

As Hino et al. (2019) suggested, Pokémon GO is useful for people as it helps them to increase their physical exercise without pain or even the intention to increase the amount that they walk; however, their study mainly focused on middle-aged and elderly people. The author team stated that many middle-aged people grew up with computer games (Generation X), so there is not much of a barrier preventing this cohort from getting familiar with Pokémon GO. The study also emphasised the 'virtuous circle' of middle-aged people playing more and walking more. Pokémon GO, a winner of the 'Sports in Life' award in Japan, has the potential to be utilised in communities to support physical exercise.

This potential contribution of Pokémon GO has been confirmed in various pieces of empirical research up to this point. Pokémon GO is a location information game that uses Google maps to search for creatures in the digital world while its players roam the real world. In line with the discussions above, it has also been found that some core fans of the game walk longer than average, and before Pokémon GO was launched, Lupton (2014) had already discussed the impact of digital technology on society's health. Various worldwide empirical analyses have discussed the fact that Pokémon GO has increased the amount of activity in the United States by 144 billion steps since it began broadcasting in July 2016, and the key factor of this contribution stems from an increase in the amount that people walk (Beach et al., 2019; Howe et al., 2016; Kogan et al., 2017; Krittanawong et al., 2017; McCartney, 2016; Watanabe et al., 2017). Other researchers have discussed in more detail the positive impact of digital gaming on mental health (Colder et al., 2018; Loveday & Burgess, 2017; Tateno et al., 2016; Watanabe et al., 2017; Zach & Tussyadiah, 2017).

2.3 Connectedness and belonging to a virtual community

As noted above, Pokémon GO can be a catalyst for increasing people's walking behaviour, as well as allowing people to communicate with other players and perhaps contributing to improving people's health and happiness through their enjoyment of the game. The entertainment of gaming could be further enhanced with the IoS and other advancements in technology soon. That is to say that there would not only be the current happiness that people gain from gaming, but by exploring Pokémon using AR and VR, players could also feel more connected with other players and could enjoy a stronger sense of community belonging (Kim et al., 2020). In line with the discussion, Dalisay et al. (2015) pointed out the civic potential of digital gaming from the perspective of nurturing social capital among players.

    Pokémon GO is now not only a game but also a catalyst for enhancing social connectedness and positive life experiences (Bonus et al., 2018; Jung, 2020; Purda & Mačak, 2019; Vella et al., 2019).

2.4 Key take-aways and analytical framework

The key take-aways from the existing literature are summarised with sub-themes in Table 1. The developed key take-aways and sub-themes are rearranged in the context of how to enhance the positive impact of digital gaming on players' well-being. To present the analytical dimensions, Figure 1 suggests analytical steps as a further research framework. Figure 1 shows the cycle of impactful relationships following the dimensions of a 'real layer' (i.e. lockdown and its limited activities) and a virtual community setting that can be accessed by joining digital games (allowing for the virtual value of connectedness, virtual collaboration with team members, and the feeling of belonging to community). Particularly in the COVID-19 lockdown situation, it could be valuable for people to implement virtual settings in their daily lives to enhance their mental health and well-being by expanding their life experiences by participating in virtual gaming with other members.

**Table 1 Key take-outs from the review**

| Key take-outs | Descriptions | Supportive academic sources |
| --- | --- | --- |
| KT1: Overview and evaluation of digital gaming | KT1-1: Gaming activities and evaluation | Howe et al., 2016; McCartney, 2016; Krittanawong et al., 2017; Kogan et al., 2017; Watanabe et al., 2017; Hino et al., 2019; BBC News, 2020; Cabinet Office, 2020; Reer and Quandt, 2020; Takaahashi, 2020 |
| | KT1-1: In the lockdown situation of COVID-19 | |

| KT2: Contribution to well-being | KT2-1: Impact of digital gaming on physical health | Johnson et al, 2016; Howe et al., 2016; McCartney, 2016; Tateno et al., 2016; Colder et al., 2018; Kogan et al., 2017; Krittanawong et al., 2017; Loveday and Burgess, 2017; Watanabe et al., 2017; Zach and Tussyadiah, 2017; NHS, 2019; Hino et al., 2019; Beach et al., 2019 |
|---|---|---|
| | KT2-2: Impact of digital gaming on mental health | |
| KT3: Connectivity with virtual community members | KT3-1: Supported feeling by connectedness | Dalisay et al., 2015; Bonus et al., 2018; Purda and Mačak, 2019; Vella et al., 2019; Kim et al., 2020 |
| | KT3-2: Feeling of being embedded in the relational networks of the virtual community | |

**Figure 1 The discussion framework on how to utilise digital gaming to support players' well-being during COVID-19 lockdown**

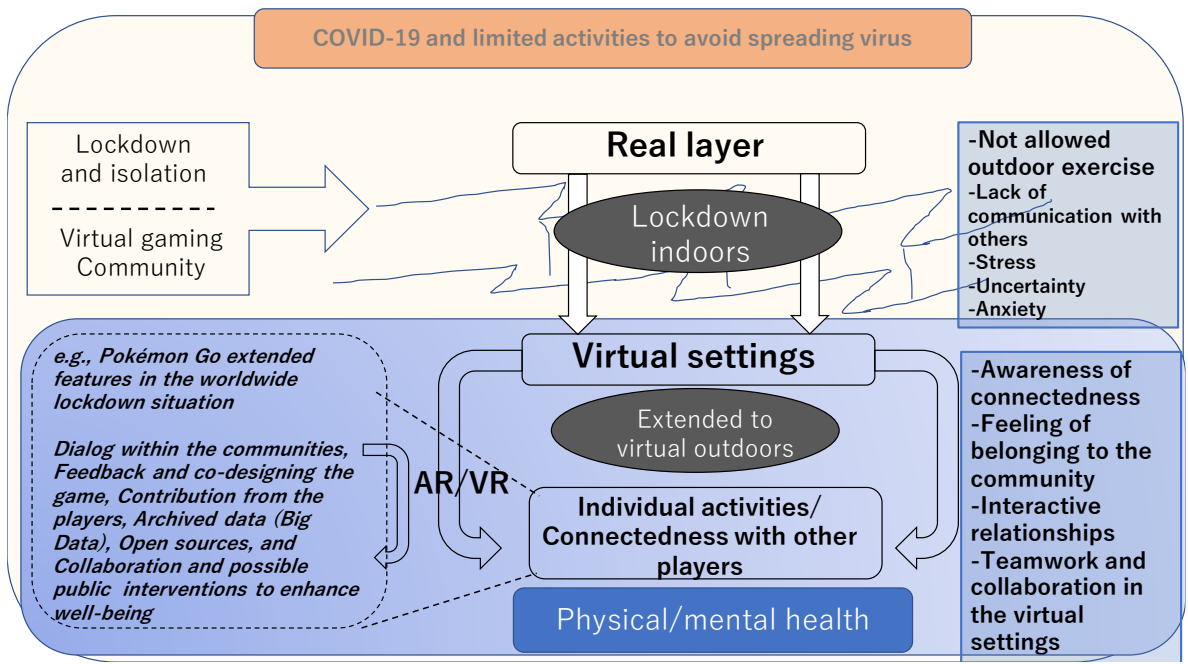

3. Contribution and further research opportunities

Digital gaming has been studied from a variety of perspectives in terms of public health: (1) its negative impact and its addictive nature; (2) its positive impact in enhancing the well-being of players by increasing their walking; (3) its positive impact in enhancing connectedness and a sense of belonging to a virtual community, which contributes to the mental health of players.

This study has narrowed down the impact of digital games and their applicability in the current difficult situation of COVID-19 lockdown. It has outlined the impact of virtual gaming on people's well-being.

As discussed above, most of the existing research has concentrated on the positive impacts of gaming on middle-aged and elderly people; there is a research gap on the impacts on the younger generation (Erikson, 1959), i.e. the 19–39 age band. This targeted age band is relevant to both Generation Z, as digital natives (James et al., 2017; Turner, 2015), and Generation Y (Venter, 2017).

Accessed April 16, 2020.

Zach FJ, Tussyadiah IP. To catch them all—the (un) intended consequences of Pokémon GO on mobility, consumption, and wellbeing. In Information and communication technologies in tourism (pp. 217-227). Springer, Cham. 2017.